\documentclass[aps,prl,preprint,groupedaddress,showpacs,amssymb,amsmath]{revtex4-1}

\newtheorem{theorem}{Theorem}

\newtheorem{remark}{Remark}
\newtheorem{example}{Example}

\begin{document}


\title{On Finite $J$-Hermitian Quantum Mechanics}


\author{Sungwook Lee}
\email[E-mail:]{sunglee@usm.edu}
\homepage[Home Page:]{http://www.math.usm.edu/lee/}
\affiliation{Department of Mathematics, University of Southern Mississippi}


\date{\today}

\begin{abstract}
In his recent paper \cite{Lee}, the author discussed $J$-Hermitian quantum mechanics and showed that $PT$-symmetric quantum mechanics is essentially $J$-Hermitian quantum mechanics. In this paper, the author discusses finite $J$-Hermitian quantum mechanics which is derived naturally from its continuum one and its relationship with finite $PT$-symmetric quantum mechanics.
\end{abstract}


\maketitle

\section{Introduction. $J$-Hermitian Quantum Mechanics}

Let $\mathcal{K}$ be the complex vector space spanned by the eigenstates of a quantum system determined by Schr\"{o}dinger equation
\begin{equation}
 i\hbar\frac{\partial\psi}{\partial t}=\hat H\psi.
\end{equation}
Due to the boundary conditions on state functions $\psi(x,t)$, without loss of generality we may assume that $\mathcal{K}$ is separable. Let $J: \mathcal{K}\longrightarrow\mathcal{K}$ be an involution i.e. $J^2=I$, the identity map. Define an inner product $\langle\ ,\ \rangle$ on $\mathcal{K}$ as follows: For any $\varphi,\psi\in\mathcal{K}$,
\begin{equation}
 \begin{aligned}
  \langle\varphi,\psi\rangle&:=\langle\varphi|J|\psi\rangle\\
  &=\int_{-\infty}^\infty\bar\varphi J\psi dx,
 \end{aligned}
\end{equation}
where $\langle\ |\ \rangle$ denotes Dirac braket.
We are interested in a particular $J=P$, the parity. That is $J$ acts on $\psi(x,t)$ as
\begin{equation}
 J\psi(x,t)=\psi(-x,t).
\end{equation}
Then the inner product is written
\begin{equation}
\label{eq:pt-product}
 \langle\varphi,\psi\rangle=\int_{-\infty}^\infty\bar\varphi(x)\psi(-x)dx.
\end{equation}
The inner product is a Hermitian product, i.e. it satisfies

\noindent IP1. $\langle\varphi,\psi\rangle=\overline{\langle\psi,\varphi\rangle}$ for any $\varphi,\psi\in\mathcal{K}$.

\noindent IP2. $\langle\varphi,a\psi_1+b\psi_2\rangle=a\langle\varphi,\psi_1\rangle+b\langle\varphi,\psi_2\rangle$ for any $\varphi,\psi_1,\psi_2\in\mathcal{K}$ and $a,b,\in\mathbb{C}$.

The squared norm $||\psi||^2=\int_{-\infty}^\infty\bar\psi(x)\psi(-x)dx$ can be positive, zero, or negative, so $\langle\ ,\ \rangle$ is an indefinite Hermitian product. The space $\mathcal{K}$ may be decomposed to
$$\mathcal{K}=\mathcal{K}^+\dotplus\mathcal{K}^-,$$
where $\mathcal{K}^+$ is spanned by the eigenstates $\psi^+_n$, $n=1,2,\cdots$ such that $\langle\psi^+_m,\psi^+_n\rangle=\delta_{mn}$ and $\mathcal{K}^-$ is spanned by the eigenstates $\psi^-_n$, $n=1,2,\cdots$ such that $\langle\psi^-_m,\psi^-_n\rangle=-\delta_{mn}$. Here, $\dotplus$ stands for orthogonal direct sum. $\mathcal{K}$ cannot be a (pre-)Hilbert space. It is called a Krein space in mathematics literature \cite{Bognar}. $J$ is required to satisfy: for each $n=1,2,\cdots$,
\begin{equation}
\label{eq:fundsymm}
\begin{aligned}
 J\psi^+_n(x,t)&=\psi^+_n(-x,t)=\psi^+_n(x,t);\\
 J\psi^-_n(x,t)&=\psi^-_n(-x,t)=-\psi^-_n(x,t).
\end{aligned}
\end{equation}
The equations \eqref{eq:fundsymm} indicate that $\psi^+_n(x,t)$ are even functions with respect to $x$, while $\psi^-_n(x,t)$ are odd functions with respect to $x$. An involution on a Krein space satisfying \eqref{eq:fundsymm} is called the \emph{fundamental symmetry} in mathematics literature \cite{Bognar}. The fundamental symmetry $J$ is used to define a positive definite inner product $\langle\ ,\ \rangle_J$ on the Krein space $\mathcal{K}$: for any $\varphi,\psi\in\mathcal{K}$,
\begin{equation}
\label{eq:j-product}
\langle\varphi,\psi\rangle_J:=\langle\varphi,J\psi\rangle=\langle\varphi|\psi\rangle.
\end{equation}
This inner product \eqref{eq:j-product} is called the $J$-inner product \cite{Bognar}. This $J$-inner product allows us to avoid awkward negative probability that was resulted by the indefinite Hermitian product \eqref{eq:pt-product}. $\mathcal{K}$ together with the $J$-inner product becomes a separable pre-Hilbert space and hence we may consider the Hilbert space of states as its completion.

Let $A: \mathcal{K}\longrightarrow\mathcal{K}$ be a bounded linear operator on a Krein space $\mathcal{K}$. Then we may define its \emph{adjoint} $A^\ast$ analogously to that in the standard Hermitian quantum mechanics. That is, $A^\ast: \mathcal{K}\longrightarrow\mathcal{K}$ is a linear operator that satisfies the property
\begin{equation}
 \langle\varphi,A\psi\rangle=\langle A^\ast\varphi,\psi\rangle
\end{equation}
for all $\varphi,\psi\in\mathcal{K}$. The adjoint satisfies the properties:

\noindent A1. $(A+B)^\ast=A^\ast+B^\ast$,

\noindent A2. $(\lambda A)^\ast=\bar\lambda A^\ast$,

\noindent A3. $(AB)^\ast=B^\ast A^\ast$,

\noindent A4. $A^{\ast\ast}=A$,

\noindent A5. if $A$ is invertible, then $(A^{-1})^\ast=(A^\ast)^{-1}$.

\noindent If we denote by $A^{[\ast]}$ the adjoint of $A$ with respect to the $J$-inner product \eqref{eq:j-product}, then $A^\ast$ and $A^{[\ast]}$ are related by (\cite{Bognar})
\begin{equation}
\label{eq:adjoint}
 JA^{[\ast]}J=A^\ast.
\end{equation}

A bounded linear operator $A: \mathcal{K}\longrightarrow\mathcal{K}$ is said to be \emph{self-adjoint} or \emph{$J$-Hermitian} or simply \emph{Hermitian} (in case there is no confusion with the notion of ordinary Hermitian operators) if $A=A^\ast$, i.e. for any $\varphi,\psi\in\mathcal{K}$
\begin{equation}
 \langle A\varphi,\psi\rangle=\langle \varphi,A\psi\rangle
\end{equation}
or
\begin{equation}
\label{eq:hermitian}
 \int_{-\infty}^\infty(\overline{A\varphi})J\psi dx=\int_{-\infty}^\infty\bar\varphi J(A\psi)dx.
\end{equation}

It turns out that a Hamiltonian of the form
\begin{equation}
\label{eq:hamiltonian}
 \hat H=-\frac{\hbar^2}{2m}\frac{\partial^2}{\partial x^2}+V(x),
\end{equation}
where the potential $V(x)$ acts on $\psi(x,t)$ by multiplication, is $J$-Hermitian if and only if it is $PT$-symmetric \cite{Lee}. Since $J$-Hermitian operators are guaranteed to have all real eigenvalues, so are $PT$-symmetric Hamiltonians. In \cite{Lee}, the author also shows that the time evolution of a state function determined by the  Hamiltonian $\hat H$ in \eqref{eq:hamiltonian} is unitary if and only if $\hat H$ is $PT$-symmetric.

In the following sections, the author will deduce the notion of $J$-Hermitian matrices as Hamiltonians and discuss the relationship between $J$-Hermitian Hamiltonians and $PT$-symmetric Hamiltonians, and their physical implications.
\section{$J$-Hermitian Matrices}
Let $\mathbb{C}^2$ denote the complex vector space
$$\mathbb{C}^2=\left\{\begin{pmatrix}
                       \alpha\\
                       \beta
                      \end{pmatrix}: \alpha,\beta\in\mathbb{C}\right\}.$$
For any $v,w\in\mathbb{C}^2$, define an inner product $\langle\ ,\ \rangle$ on $\mathbb{C}^2$ by
\begin{equation}
\label{eq:pt-product2}
\begin{aligned}
  \langle v,w\rangle&:=\langle v|J|w\rangle\\
  &=v^\dagger Jw,
  \end{aligned}
\end{equation}
where $v^\dagger=\bar v^t$. Here $\langle v|w \rangle$ stands for Dirac braket which is the standard positive definite Hermitian product $v^\dagger w$ and $J:\mathbb{C}^2\longrightarrow\mathbb{C}^2$ is an involution i.e. $J^2=I$, the identity transformation. Either $J=I$ or, up to diagonalisation and sign,
\begin{equation}
\label{eq:fundsymm2}
 J=\begin{pmatrix}
    1 & 0\\
    0 & -1
   \end{pmatrix},
\end{equation}
since $J$ has eigenvalues $\pm 1$. If $J=I$, then \eqref{eq:pt-product2} is simply Dirac braket. In this paper, we consider $J$ in \eqref{eq:fundsymm2}. 
 Then $\langle\ ,\ \rangle$ is an indefinite Hermitian product on $\mathbb{C}^2$. Carl M. Bender (for example \cite{Bender}, \cite{Bender2}) introduced the inner product \eqref{eq:pt-product2} as $(PTv)^tw$ where $P$ and $T$ stand for parity and time-reversal operator, respectively. Parity $P$ is defined to be $J$ in \eqref{eq:fundsymm2}. Carl M. Bender defined $P$ by $P=\begin{pmatrix}                                                                                                                                                                                                                                                                                                                                                                                   0 & 1\\
1 & 0                                                                                                                                                                                                                                                                                                                                                                                         \end{pmatrix}$. Since $P$ has eigenvalues $\pm 1$, $J$ is the diagonalisation of the parity used by Bender. Time-reversal operator $T$ is defined as complex conjugation. Let $$e_1=\begin{pmatrix}
   1\\
   0                                                                                                                             \end{pmatrix},\ e_2=\begin{pmatrix}
   0\\
   1
   \end{pmatrix}.$$
   Then
   $$\langle e_1,e_1\rangle=1,\ \langle e_1,e_2\rangle=0,\ \langle e_2,e_2\rangle=-1.$$
   Thus, $\{e_1,e_2\}$ is an orthonormal basis of $\mathbb{C}^2$ with respect to the indefinite Hermitian product \eqref{eq:pt-product2}. The involution $J$ \eqref{eq:fundsymm2} satisfies
   \begin{equation}
    \begin{aligned}
     Je_1&=e_1,\\
     Je_2&=-e_2.
    \end{aligned}
   \end{equation}
So, $J$ is the fundamental symmetry and $(\mathbb{C}^2,J)$ is a 2-dimensional Krein space. Let us define another inner product $\langle\ ,\ \rangle_J$ as follows. For any $v,w\in\mathbb{C}^2$,
\begin{equation}
\begin{aligned}
\label{eq:j-product2}
 \langle v,w\rangle_J:=\langle v,Jw\rangle.
\end{aligned}
\end{equation}
Then $\langle v,w\rangle_J=\langle v|w\rangle=v^\dagger w$ i.e. the usual Dirac braket. We call $\langle\ ,\ \rangle_J$ \emph{$J$-inner product}. $\mathbb{C}^2$ with $J$-inner product is a 2-dimensional Hilbert space.

Let $\hat H:\mathbb{C}^2\longrightarrow\mathbb{C}^2$ be a linear operator on $\mathbb{C}^2$. We find the matrix representation of the adjoint $\hat H^\ast$. First write the matrix representation of $\hat H$ as $\begin{pmatrix}
             a & b\\
c & d
            \end{pmatrix}$ where $a,b,c,d\in\mathbb{C}$. Then
\begin{align*}
 \hat He_1&=ae_1+be_2,\\
 \hat He_2&=ce_1+de_2.
\end{align*}
So, we obtain
\begin{align*}
 a&=\langle e_1,\hat He_1\rangle,\ b=-\langle e_2,\hat He_1\rangle,\\
 c&=\langle e_1,\hat He_2\rangle,\ d=-\langle e_2,\hat He_2\rangle.
\end{align*}
Let $\begin{pmatrix}
             a^\ast & b^\ast\\
c^\ast & d^\ast
            \end{pmatrix}$ be the matrix representation of the adjoint $\hat H^\ast$. Then
            \begin{align*}
             \hat H^\ast e_1&=a^\ast e_1+b^\ast e_2,\\
 \hat H^\ast e_2&=c^\ast e_1+d^\ast e_2.
            \end{align*}
So, we obtain
\begin{align*}
 a^\ast&=\langle e_1,\hat H^\ast e_1\rangle,\ b^\ast=-\langle e_2,\hat H^\ast e_1\rangle,\\
 c^\ast&=\langle\langle e_1,\hat H^\ast e_2\rangle,\ d^\ast=-\langle e_2,\hat H^\ast e_2\rangle.
\end{align*}
Now,
\begin{align*}
a^\ast&=\langle e_1,\hat H^\ast e_1\rangle=\langle \hat He_1,e_1\rangle=\overline{\langle e_1,\hat He_1\rangle}=\bar a,\\
b^\ast&=-\langle e_2,\hat H^\ast e_1\rangle=-\langle\hat He_2,e_1\rangle=-\overline{\langle e_1,\hat He_2\rangle}=-\bar c,\\
c^\ast&=\langle e_1,\hat H^\ast e_2\rangle=\langle\hat He_1,e_2\rangle=\overline{\langle e_2,\hat He_1\rangle,}=-\bar b,\\
d&=-\langle e_2,\hat He_2\rangle=-\langle\hat He_2,e_2\rangle=-\overline{\langle e_2,\hat He_2\rangle}=\bar d.
\end{align*}
Thus, the matrix representation of the adjoint $\hat H^\ast$ is given by
\begin{equation}
 \hat H^\ast=\begin{pmatrix}
           \bar a & -\bar c\\
-\bar b & \bar d
             \end{pmatrix}.
\end{equation}
If $\hat H$ is $J$-Hermitian i.e. $\hat H=\hat H^\ast$, then $\hat H$ is 
\begin{equation}
\label{eq:hermitianmatrix}
 \hat H=\begin{pmatrix}
         a & b\\
-\bar b & d
        \end{pmatrix},
\end{equation}
where $a$ and $d$ are real. We call a matrix of the form \eqref{eq:hermitianmatrix} a $J$-Hermitian matrix. $2\times 2$ $J$-Hermitian matrices can be characterised as follows.
\begin{theorem}
 A $2\times 2$ matrix $\hat H$ is $J$-Hermitian if and only if 
\begin{equation}
 \label{eq:hermitianmatrix2}
J\hat H^\dagger J=\hat H.
\end{equation}
\end{theorem}
\eqref{eq:hermitianmatrix2} is indeed identical to \eqref{eq:adjoint}.
In general, we have
\begin{theorem}
 An $n\times n$ matrix $\hat H$ is $J$-Hermitian if and only if
\begin{equation}
 J\hat H^\dagger J=\hat H,
\end{equation}
where
$$J=\begin{pmatrix}
         1 & & & & & &\\
&  & \ddots & & & &\\
& & & & & 1 &\\
& & & & & & -1
        \end{pmatrix}.$$
\end{theorem}
\begin{example}
 If a linear operator $\hat H:\mathbb{C}^3\longrightarrow\mathbb{C}^3$ is $J$-Hermitian, then $\hat H$ may be written as
\begin{equation}
\hat H=\begin{pmatrix}
        a & b & c\\
\bar b & d & e\\
-\bar c & -\bar e & f
       \end{pmatrix}
\end{equation}
where $a$, $d$, and $f$ are real.
\end{example}
\section{Time Evolution}
Let $\hat H$ be a $2\times 2$ complex matrix. The Schr\"{o}dinger equation
\begin{equation}
i\hbar\frac{d\psi(t)}{dt}=\hat H\psi(t)  
\end{equation}
has solution
\begin{equation}
 \psi(t)=\hat U(t)\psi(0)
\end{equation}
where
\begin{equation}
\label{eq:time-evolution}
\hat U(t)=\exp\left(-\frac{i}{\hbar}\hat Ht\right).
\end{equation}
Suppose that $\langle\psi(0),\psi(0)\rangle\ne 0$. Physically we want $\hat U(t)$ to be unitary, i.e.
\begin{align*}
 \langle\psi(t),\psi(t)\rangle&=\langle \hat U(t)\psi(0),\hat U(t)\psi(0)\rangle\\
&=(\hat U(t)\psi(0))^\dagger J\hat U(t)\psi(0)\\
&=\psi^\dagger(0)\hat U^\dagger(t)J\hat U(t)\psi(0)\\
&=\psi^\dagger(0)J\psi(0)\\
&=\langle\psi(0),\psi(0)\rangle.
\end{align*}
Since $\langle\psi(0),\psi(0)\rangle\ne 0$, we see that $\hat U(t)$ is unitary if and only if 
\begin{equation}
\label{eq:unitary}
 \hat U^\dagger(t)J\hat U(t)=J.
\end{equation}
The set of unitary transformations i.e. transformations of $\mathbb{C}^2$ that satisfy \eqref{eq:unitary} forms a Lie subgroup of $SL(2,\mathbb{C})$ which is the well-known indefinite unitary group $U(1,1)$. This is consistent with the fact that the structure group of the frame bundle $L\mathbb{C}^2$ is $U(1,1)$ when $\mathbb{C}^2$ is considered as an indefinite 2-dimensional Hermitian manifold. Since $J=J^{-1}$,
\begin{equation}
\label{eq:unitary2}
 J\hat U^\dagger(t)J\hat U(t)=\exp\left(\frac{i}{\hbar}J\hat H^\dagger Jt\right)\exp\left(-\frac{i}{\hbar}\hat Ht\right).
\end{equation}
Clearly, if $J\hat H^\dagger J=\hat H$ then $\hat U(t)$ is unitary. Conversely, if $\hat U(t)$ is unitary then by differentiating \eqref{eq:unitary2} at $t=0$, we obtain $J\hat H^\dagger J=\hat H$. Thus, we have
\begin{theorem}
 $\hat U(t)$ in \eqref{eq:time-evolution} is unitary if and only if $\hat H$ is $J$-Hermitian.
\end{theorem}
\begin{remark}
 A $2\times 2$ matrix $\hat H$ is $J$-Hermitian if and only if $-i\hat H\in\mathfrak{u}(1,1)$, the Lie algebra of $U(1,1)$. Any indefinite Hermitian manifold is orientable so the structure group of the frame bundle $L\mathbb{C}^2$ may be reduced to the special indefinite unitary group $SU(1,1)$. The Lie algebra $\mathfrak{su}(1,1)$ is the set of elements in $\mathfrak{u}(1,1)$ that are trace-free. With the additional condition $\mathrm{tr}(\hat H)=0$, \eqref{eq:hermitianmatrix} may be written as
 \begin{equation}
  \label{eq:hermitianmatrix3}
  \hat H=\begin{pmatrix}
          a & b\\
          -\bar b & -a
         \end{pmatrix}
 \end{equation}
where $a$ is real.
\end{remark}
\section{The Reality of Eigenvalues}
Let $\hat H$ be a $J$-Hermitian matrix and $E$ an eigenvalue of $\hat H$. Suppose that $E$ is a non-real complex number $E=\alpha+i\beta$ where $\alpha$ and $\beta\ne 0$ are real. Let $v$ be an eigenvector of $\hat H$ with the eigenvalue $E$. Then $\hat Hv=Ev$. Multiply this equation by $\bar E$. Then we get $\bar E\hat Hv=|E|^2v$ or $\frac{\bar E}{|E|^2}\hat Hv=v$. Now,
\begin{align*}
 ||v||^2&=v^\dagger Jv\\
 &=\left[\frac{\bar E}{|E|^2}\hat Hv\right]^\dagger Jv\\
 &=\frac{E}{|E|^2}v^\dagger\hat H^\dagger Jv\\
 &=\frac{E}{|E|^2}v^\dagger J\hat Hv\\
 &=\frac{E^2}{|E|^2}||v||^2.
\end{align*}
From this we obtain $(E^2-|E|^2)||v||^2=0$. Since $E$ is a non-real complex number, $E^2\ne |E|^2$ and hence $||v||^2=0$. Note that the inner product is indefinite, so $||v||^2=0$ does not mean that $v=O$. Hence, the eigenvalues of a $J$-Hermitian matrix are all real as long as the eigenvectors they are belonging to have non-vanishing squared norms. Furthermore, an eigenvector with zero eigenvalue must have vanishing norm. To see this, let $v=(\alpha,\beta)^t$ be an eigenvector with zero eigenvalue. Then the equation $\hat Hv=O$ may be written as
\begin{align}
\label{eq:zeroeigenvalue}
 a\alpha+b\beta&=0,\\
 \label{eq:zeroeigenvalue2}
 -\bar b\alpha-a\beta&=0.
\end{align}
If \eqref{eq:zeroeigenvalue} and \eqref{eq:zeroeigenvalue2} are not equivalent or $\det\hat H\ne 0$, then $v=(0,0)^t$. If \eqref{eq:zeroeigenvalue} and \eqref{eq:zeroeigenvalue2} are equivalent or $\det\hat H=0$, then $v$ may be chosen to be $v=(-b,a)^t$. Now,
\begin{align*}
 ||v||^2&=\begin{pmatrix}
           -\bar b & \bar a
          \end{pmatrix}
\begin{pmatrix}
                           1 & 0\\
                           0 & -1
                          \end{pmatrix}\begin{pmatrix}
                          -b\\
                          a
                          \end{pmatrix}\\
                          &=|b|^2-a^2\\
                          &=\det\hat H\\
                          &=0.
\end{align*}
The characteristic equation $\det(\hat H-EI)=0$ is equivalent to $E^2=a^2-|b|^2$ where $E$ is an eigenvalue of $\hat H$. We require that each eigenvalue $E$ is non-zero real, so the $J$-Hermitian Hamiltonian $\hat H$ must satisfy
\begin{equation}
\label{eq:realeigenvalue}
 \det\hat H<0.
\end{equation}
$\hat H$ has two distinct real eigenvalues, one positive and the other negative
\begin{equation}
\label{eq:eigenvalue}
 E_{\pm}=\pm\sqrt{a^2-|b|^2}.
\end{equation}
Let $v_{\pm}$ denote eigenvectors with eigenvalues $E_{\pm}$, respectively. The equation $\hat Hv=Ev$ is written as
\begin{align}
\label{eq:eigenequation1}
 (a-E)\alpha+b\beta&=0\\
 \label{eq:eigenequation2}
 -\bar b\alpha-(a+E)\beta&=0
\end{align}
where $v=(\alpha,\beta)^t$. Without loss of generality we may assume that $a>0$. Denote by $v_+$ and $v_-$ eigenvectors with eigenvalues $E_+$ and $E_-$, respectively. From the equation \eqref{eq:eigenequation1}, $v_+$ may be chosen to be $v_+=(-b,a-E_+)^t$. Then
\begin{eqnarray*}
 ||v_+||^2&=\begin{pmatrix}
           -\bar b & a-E_+
          \end{pmatrix}\begin{pmatrix}
          1 & 0\\
          0 & -1
          \end{pmatrix}\begin{pmatrix}
          -b\\
          a-E_+
          \end{pmatrix}\\
          &=|b|^2-(a-E_+)^2\\
          &=-2E_+(E_+-a).
\end{eqnarray*}
From the equation \eqref{eq:eigenequation2}, $v_-$ may be chosen to be $v_-=(a+E_-,-\bar b)^t$. Then
\begin{eqnarray*}
 ||v_-||^2&=\begin{pmatrix}
           a+E_- & -b
          \end{pmatrix}\begin{pmatrix}
          1 & 0\\
          0 & -1
          \end{pmatrix}\begin{pmatrix}
          a+E_-\\
          -\bar b
          \end{pmatrix}\\
          &=(a+E_-)^2-|b|^2\\
          &=2E_-(E_-+a).
\end{eqnarray*}
From \eqref{eq:eigenvalue} we find that $E_+-a<0$ and $E_-+a>0$. Thus $v_+$ has a positive squared norm and $v_-$ has a negative squared norm. Moreover, $v_+$ and $v_-$ are orthogonal to each other:
\begin{align*}
 \langle v_+,v_-\rangle&=\begin{pmatrix}
                          -\bar b & a-E_+
                         \end{pmatrix}\begin{pmatrix}
                         1 & 0\\
                         0 & -1
                         \end{pmatrix}\begin{pmatrix}
                                       a+E_-\\
                                       -\bar b
                                      \end{pmatrix}\\
                                      &=\begin{pmatrix}
                          -\bar b & -(a-E_+)
                         \end{pmatrix}\begin{pmatrix}
                                       a+E_-\\
                                       -\bar b
                                      \end{pmatrix}\\
                                      &=-\bar b(a+E_-)+\bar b(a-E_+)\\
                                      &=0.
                                      \end{align*}
The zero value in the last line is obtained by \eqref{eq:eigenvalue}.
\section{$J$-Hermitian and $PT$-symmetric Hamiltonians}
In this section, we study the relationship between $J$-Hermitian and $PT$-symmetric Hamiltonians.

Let $\hat H$ be a $2\times 2$ complex matrix. The parity $P=J$ acts on $\hat H$ as
$P\hat H P^{-1}$. As mentioned earlier, time-reversal operator $T$ is defined to be complex conjugation. $\hat H$ is said to be $PT$-symmetric if
\begin{equation}
 P\overline{\hat H} P^{-1}=\hat H.
\end{equation}
Suppose that $\hat H$ is $PT$-symmetric. In addition, we must require that $\hat H$ is symmetric. Otherwise, time evolution determined by the $PT$-symmetric Hamiltonian would not be unitary. Then $\hat H$ is of the form 
$$\hat H=\begin{pmatrix}
                                  a & ib\\
ib & c
                                 \end{pmatrix},$$
where $a,b,c$ are real numbers. Thus, $\hat H$ is also $J$-Hermitian. $PT$-symmetric Hamiltonians are a special class of $J$-Hermitian Hamiltonians.
\section{Time Evolution with $J$-Inner Product}
In $J$-Hermitian quantum mechanics, negative probabilities are merely an artifact of the indefinite Hermitian inner product \eqref{eq:pt-product2}, and there appears to be no physically meaningful notion for negative probabilities. So one may argue that we must use $J$-inner product \eqref{eq:j-product2} for doing physics with $J$-Hermitian quantum mechanics. Since $J$-inner product is the same as the usual Dirac braket, time evolution operator \eqref{eq:time-evolution} is unitary with respect to $J$-inner product if and only if Hamiltonian is Hermitian in ordinary sense i.e. $\hat H^\dagger=\hat H$. Thus, in order for time evolution determined by a $J$-Hermitian Hamiltonian $\hat H$ to be unitary with respect to $J$-inner product, it is also required to be Hermitian in ordinary sense which results $\hat H$ in the form
\begin{equation}
\hat H=\begin{pmatrix}
        a & 0\\
        0 & -a
       \end{pmatrix}, 
\end{equation}
i.e. a real diagonal matrix. Clearly $\pm a$ are the eigenvalues of $\hat H$.

On the other hand, since a $J$-Hermitian matrix $\hat H$ has 2 distinct real eigenvalues, it is always diagonalisable i.e. there exists a $2\times 2$ invertible matrix $\hat M$ such that
$$\hat M^{-1}\hat H\hat M=\begin{pmatrix}
                 \lambda_1 & 0\\
                 0 & \lambda_2
                \end{pmatrix},$$
where $\lambda_1,\lambda_2$ are the two distinct real eigenvalues of $\hat H$. In fact, the matrix $\hat M$ is given by the block matrix of its column vectors $v_1,v_2$
$$\hat M=\begin{pmatrix}
     v_1 & v_2
    \end{pmatrix},$$
where $v_1,v_2$ are eigenvectors with the eigenvalues $\lambda_1,\lambda_2$, respectively. If $v_1,v_2$ are orthonormal vectors such that $\langle v_1,v_1\rangle=1$, $\langle v_1,v_2\rangle=0$, and $\langle v_2,v_2\rangle=-1$, then $\hat M^\dagger J\hat M=J$ i.e. $\hat M$ is unitary. If $\hat H$ is $J$-Hermitian, then for any unitary matrix $\hat M$, $\hat U(t)=\exp\left(-\frac{i}{\hbar}\hat M^{-1}\hat H\hat Mt\right)$ is unitary and so $\hat M^{-1}\hat H\hat M$ is $J$-Hermitian.
    
Both Hamiltonians $\hat H$ and its diagonalisation $M^{-1}\hat H M$ have the same eigenvalues (observables) and physically the quantum mechanical systems from the two Hamiltonians are not distinguishable. Therefore, as long as time-independent Hamiltonians are concerned, finite $J$-Hermitian or equivalently $PT$-symmetric quantum mechanics is essentially the standard Hermitian quantum mechanics.
\section{The Symmetry of $J$-Hermitian Quantum Mechanics}
As seen earlier, the set of unitary transformations of the 2-dimensional Krein space $\mathbb{C}^2$ forms indefinite unitary group $U(1,1)$. Hence, $U(1,1)$ is the symmetry group of the 2-dimensional Krein space $\mathbb{C}^2$. As discussed in the preceding section, time evolution is also required to be unitary with respect to $J$-inner product which is the usual Dirac braket. The symmetry group of $\mathbb{C}^2$ as a 2-dimensional Hilbert space is $U(2)$. Thus, $J$-Hermitian quantum mechanics must have $U(1,1)\cap U(2)$ symmetry. Suppose that $U\in U(1,1)\cap U(2)$. Then
\begin{equation}
\label{eq:j-unitary}
U^\dagger\begin{pmatrix}
            1 & 0\\
            0 & -1
           \end{pmatrix}U=\begin{pmatrix}
            1 & 0\\
            0 & -1
           \end{pmatrix}
           \end{equation}
           and
           \begin{equation}
            U^\dagger U=I.
           \end{equation}
           Multiplying \eqref{eq:j-unitary} by $U$ from the left, we obtain
           \begin{equation}
           \label{eq:torus}
            \begin{pmatrix}
             1 & 0\\
             0 & -1
            \end{pmatrix}U=U\begin{pmatrix}
            1 & 0\\
            0 & -1
            \end{pmatrix}.
           \end{equation}
Let $U=\begin{pmatrix}
        a & b\\
        c & d
       \end{pmatrix}$ where $a,b,c,d\in\mathbb{C}$. Then by \eqref{eq:torus}, we find that $b=c=0$ and so
       \begin{equation}
       \label{eq:torus2}
        U=\begin{pmatrix}
            a & 0\\
            0 & d
           \end{pmatrix}.
       \end{equation}
           Since $U^\dagger U=I$, $a$ and $d$ satisfy
           \begin{equation}
           \label{eq:circle}
            |a|^2=|d|^2=1.
           \end{equation}
           Conversely, any complex matrix $U$ of the form \eqref{eq:torus2} satisfying \eqref{eq:circle} is in the intersection $U(1,1)\cap U(2)$. Hence,
           \begin{align*}
           U(1,1)\cap U(2)&=\left\{\begin{pmatrix}
                e^{i\theta} & 0\\
                0 & e^{i\phi}
               \end{pmatrix}: 0\leq \theta,\phi\leq 2\pi\right\}\\
               &\cong \{(e^{i\theta},e^{i\phi}): 0\leq \theta,\phi\leq 2\pi\}\\
               &=SO(2)\times SO(2).
               \end{align*}
That is, $J$-Hermitian quantum mechanics has toroidal symmetry. When $\mathbb{C}^2$ is considered as a 2-dimensional Hermitian or indefinite Hermitian manifold, the symmetry group of $\mathbb{C}^2$ as a 2-dimensional Hilbert or Krein space is the same as the structure group of the frame bundle $L\mathbb{C}^2$. Since a Hermitian manifold is orientable, the structure group may be reduced to $SU(2)$ or $SU(1,1)$ depending on the signature of $\mathbb{C}^2$. As a result, the symmetry of $J$-Hermitian quantum mechanics may be reduced to $SU(1,1)\cap SU(2)=SO(2)$. Although finite $J$-Hermitian or $PT$-symmetric quantum is essentially the standard Hermitian quantum mechanics, geometrically it exhibits a distinct symmetry.
\section{Conclusion}
The author derived the notion of $J$-Hermitian matrices naturally from the notion of $J$-Hermitian Hamiltonians he discussed in \cite{Lee}. $J$-Hermitian matrices may serve as Hamiltonians for finite dimensional quantum mechanical systems. When $\mathbb{C}^2$ is equipped with standard indefinite Hermitian product \eqref{eq:pt-product2} with $J$ in \eqref{eq:fundsymm2}, time evolution of a state vector determined by a Hamiltonian is unitary if and only if the Hamiltonian is $J$-Hermitian. It turns out that a $PT$-symmetric Hamiltonian $\hat H$ is $J$-Hermitian if $\hat H$ is also symmetric. Note that in continuum case \cite{Lee}, a $PT$-symmetric Hamiltonian is automatically symmetric because it is self-adjoint. The author argued that finite  $J$-Hermitian or $PT$-symmetric quantum mechanics is essentially the standard Hermitian quantum mechanics as long as time-independent Hamiltonians are concerned. Hence there is no issue with unitarity of time evolution with $J$-inner product unlike its continuum case \cite{Lee}.

The author's colleague and a physicist Lawrence R. Mead observed the violation of unitarity of a time evolution with $J$-inner product when concerned $J$-Hermitian or $PT$-symmetric Hamiltonian is time-dependent \cite{Mead}. This is perhaps due to the fact that time-dependent $J$-Hermitian Hamiltonian $\hat H(t)$ is not diagonalisable except when $t=0$. Therefore, the only physically acceptable time-dependent $J$-Hermitian Hamiltonians appear to be the diagonal ones. This issue will be investigated further.

\end{document}